\documentclass[preprint,12pt]{elsarticle}
\usepackage{graphicx}
\begin{document}
\begin{frontmatter}

\title{Coupled structural and magnetic properties of ferric fluoride nanostructures: part II, a Monte-Carlo Heisenberg study}
\author[a,b]{B.  Fongang} 
\author[a]{Y. Labaye}
\author[a]{F. Calvayrac}
\ead{Florent.Calvayrac@univ-lemans.fr}
\author[b]{S. Zekeng}
\author[a]{J.M. Gren\`eche}
\address[a]{
Laboratoire de Physique de l'Etat Condens\'e, CNRS UMR 6087, IRIM2F CNRS FR 2575
Universit\'e du Maine, 72085 Le Mans, Cedex 9, France}
\address[b]{
Laboratoire de Sciences des Mat\'eriaux, D\'epartement de Physique, 
Universit\'e de Yaound\'e I, BP 812 Yaound\'e, Cameroun
}

\begin{abstract}

We present a numerical study of the magnetic structure of nanostructured
iron fluoride, using the Monte-Carlo-Metropolis simulated annealing
technique and a classical Heisenberg Hamiltonian with a superexchange
angle dependence. The parameters are adjusted on experimental results,
and the atomic structure and topology taken from a previous atomistic
model of grain boundaries in the same system.  We find perfect
antiferromagnetic crystalline grains and a disordered magnetic configuration
(speromagnetic like) at the grain boundary, in agreement with experimental
findings. Both the lowest magnetic energy and the rate of magnetic
frustration are found to be dependent on the relative disorientation
of crystalline grains, i.e. on the cationic topology. By simulating
hysteresis loops, we find that the magnetization
rotation is not spatially uniform.  We conclude on possible
extensions of the model.

\end{abstract}


\end{frontmatter}

\section{Introduction} During the last decade, strong efforts have been
devoted  studying the magnetic properties of nanostructures.  This
growing interest is due to their unusual and tunable
physical properties, which are strongly influenced by the effect of
confinement resulting in a large contribution of surface and interface
effects
\cite{Gleiter}.  Understanding the correlation between magnetic
properties and nanostructure involves collaborative efforts between
chemists, physicists and materials scientists to study both fundamental
properties and potential applications
\cite{Hernando}.  This can be done experimentally using diffraction
techniques, local probe techniques  such
as M\"ossbauer spectrometry, nuclear magnetic resonance, electron
microscopies and magnetic measurements. 
In addition,  one can use numerical modeling  techniques based 
on {\em ab initio}  or phenomenological calculations such as the
classical Monte Carlo methods with the Heisenberg model, as in 
\cite{Crisan,Greneche1,Miglierini,Miglierini2} for instance.  The key issue for
understanding  magnetic macroscopic properties, such as magnetization
or susceptibility, would be to investigate the contributions arising
from the nanograins and the role played by the nanograin surface and the
interface between grains generally named grain boundary (GB).  GBs in
magnetic nanostructures feature a disordered atomic structure and
a spin-glass-like behavior and have a chemical composition which might differ
strongly from that of the corresponding nanograin
\cite{Greneche2,Suryanarayana,Coey}.\\

To investigate the magnetic behavior of  nanostructures and
the mutual  influence of the grains and grain boundaries,
 we selected ferric
fluoride ($ FeF_3 $), because this ionic phase 
displays several advantages. Indeed, it  
exhibits a fascinating polymorphism with three different crystalline phases and
two amorphous varieties, the structures of which result from ordered and
disordered packing of corner-sharing octahedral units respectively.
Those rather different cationic topologies originate collinear and
non-collinear magnetic arrangements in conjunction with the
antiferromagnetic nature of the superexchange interactions. Iron
fluoride is therefore considered as an excellent simple model system
to illustrate the concept of topological magnetic frustration.  The more
stable crystalline form of $ FeF_3 $, which is the rhombohedral phase (r-$ FeF_3
$; SG $ R\bar{3}C $) is antiferromagnetically ordered below $ T_N $ =
363K while the amorphous varieties display a speromagnetic behavior
 below $ T_F $~30-40K
\cite{Ferey,Ferey2,Ferey3}. The presence of fluorine in those materials
establishes superexchange magnetic interactions between
magnetic ions.  The strength of the coupling is linearly dependent on $
\cos^2(\theta) $, where $ \theta $ denotes the 
superexchange angle $Fe-F-Fe $ 
while the sign is given by the orbital configuration of the $ FeFFe $
bond:  for example, $ 180^\circ $ ($ Fe^{3+}-F^--Fe^{3+} $) is
characterized by a strong antiferromagnetic coupling according to
Kanamori-Goodenough rules
\cite{kana,Good}.  \\
The magnetic properties of $ FeF_3 $ have been intensively studied during
the past two decades and can be described in terms of iron ring
statistics with induced magnetic frustration
\cite{Greneche2,Greneche3}.\\
\begin{itemize}
    \item
        the rhombohedral phase r-$ FeF_3 $, which is a pseudo-cubic
        packing of weakly tilted octahedral $ FeF_6 $ units, is not
        frustrated;
    \item
        the hexagonal tungsten bronze HTB-$ FeF_3 $ which results from
        the superposition of magnetically frustrated
	planes composed of hexagonal and triangular
        octahedral units cycles ;
    \item
        and the pyrochlore pyr-$ FeF_3 $ phase which consists of a
        packing of octahedral units building corner-sharing tetrahedron  is thus
        more frustrated than HTB-$ FeF_3 $.
\end{itemize}
Recent experimental and theoretical  studies on the 
properties of $ FeF_3 $ nanostructures obtained by
mechanical milling have shown that, at the nanoscale, this material is
composed of two parts:  the grain which behaves as the crystalline r-$
FeF_3 $ phase and a disordered grain boundary which remains composed of
corner-sharing octahedral units
\cite{Guerault,Fongang}.  In terms of ring statistics, it has been
numerically established that all odd rings 
are localized in the grain boundary of nanostructures, confirming
the speromagnetic structure evidenced from in-field M\"ossbauer 
spectrometry.
The Monte Carlo-Metropolis  method has been successfully applied  
 to  study the
surface and finite-size effects in nanostructures
\cite{Berger,Kachkachi,Iglesias}.  Nevertheless, no theoretical studies
on the magnetic properties of ionic nanostructures which focus 
on the role of grain boundary by Monte Carlo simulation have yet been
reported, to our knowledge.  Unlike micromagnetic or molecular field
calculations, the Monte Carlo simulation can take into account the
atomic structure of the lattice and the nature of superexchange
interactions for ionic materials such as $ FeF_3 $.\\
The Monte Carlo-Metropolis simulated annealing 
technique is an effective approach in the study
of a system with many degrees of freedom.  During such a simulation,
random numbers are used to 
sample conformations of the system  with 
correct thermodynamical probabilities.  A
typical Monte Carlo simulation consists of two steps:  thermalization
and sampling.  During thermalization, the system is led adiabatically
to its thermodynamical equilibrium.  After the system reaches this
equilibrium,  properties of
interest can be estimated by averaging over enough samples
\cite{Binder,Newman}. Besides, in the simulated annealing
method, it is possible to find heuristically the lowest
energy state of the system. \\
We used, in this work, the standard Metropolis algorithm to simulate the
magnetic behavior of iron fluoride nanostructure with an emphasis on
the role of grain boundaries.
\section{Magnetic model} 
\begin{figure*}[!ht]
    \centering
     \includegraphics[scale=0.15]{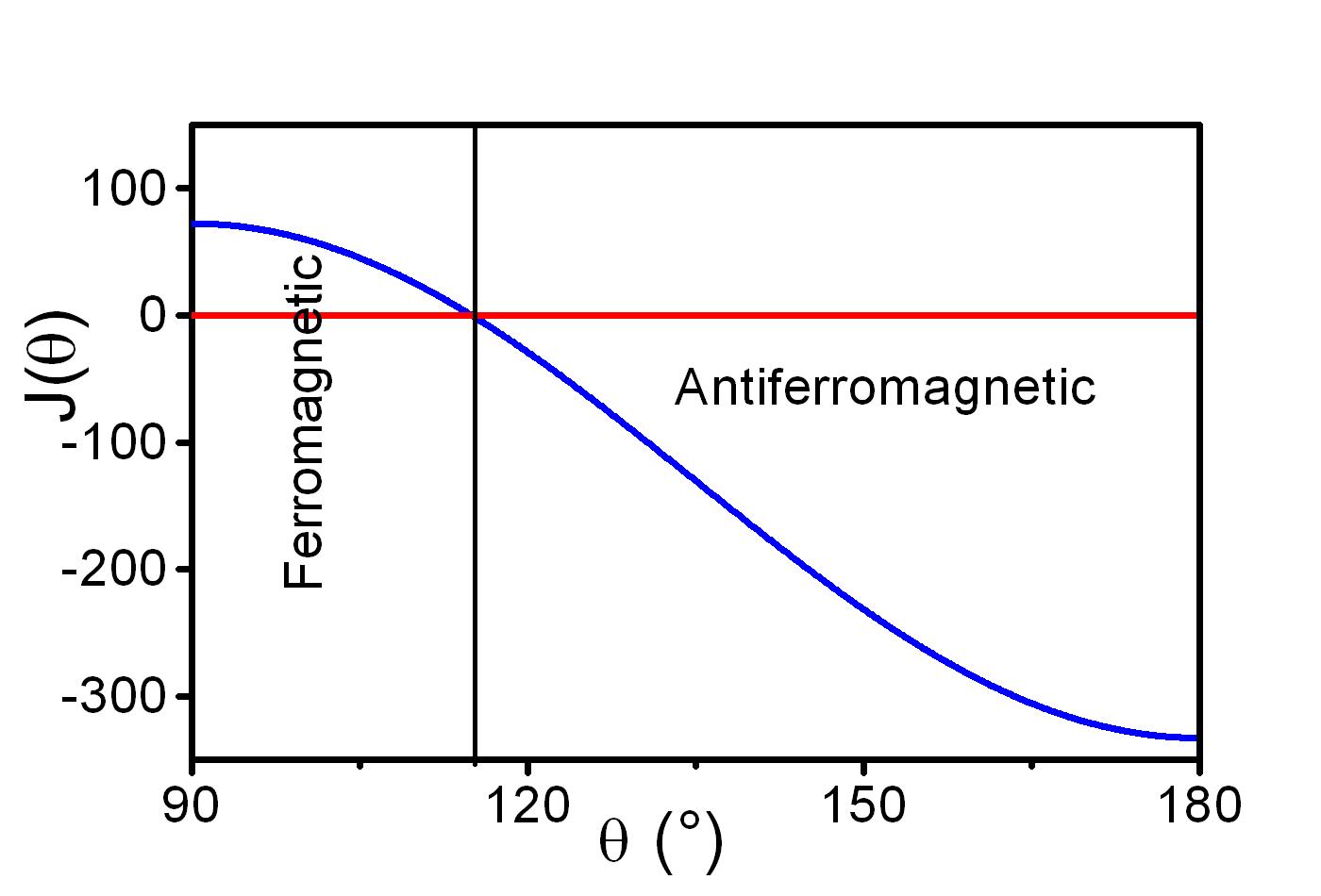}
     \includegraphics[scale=0.15]{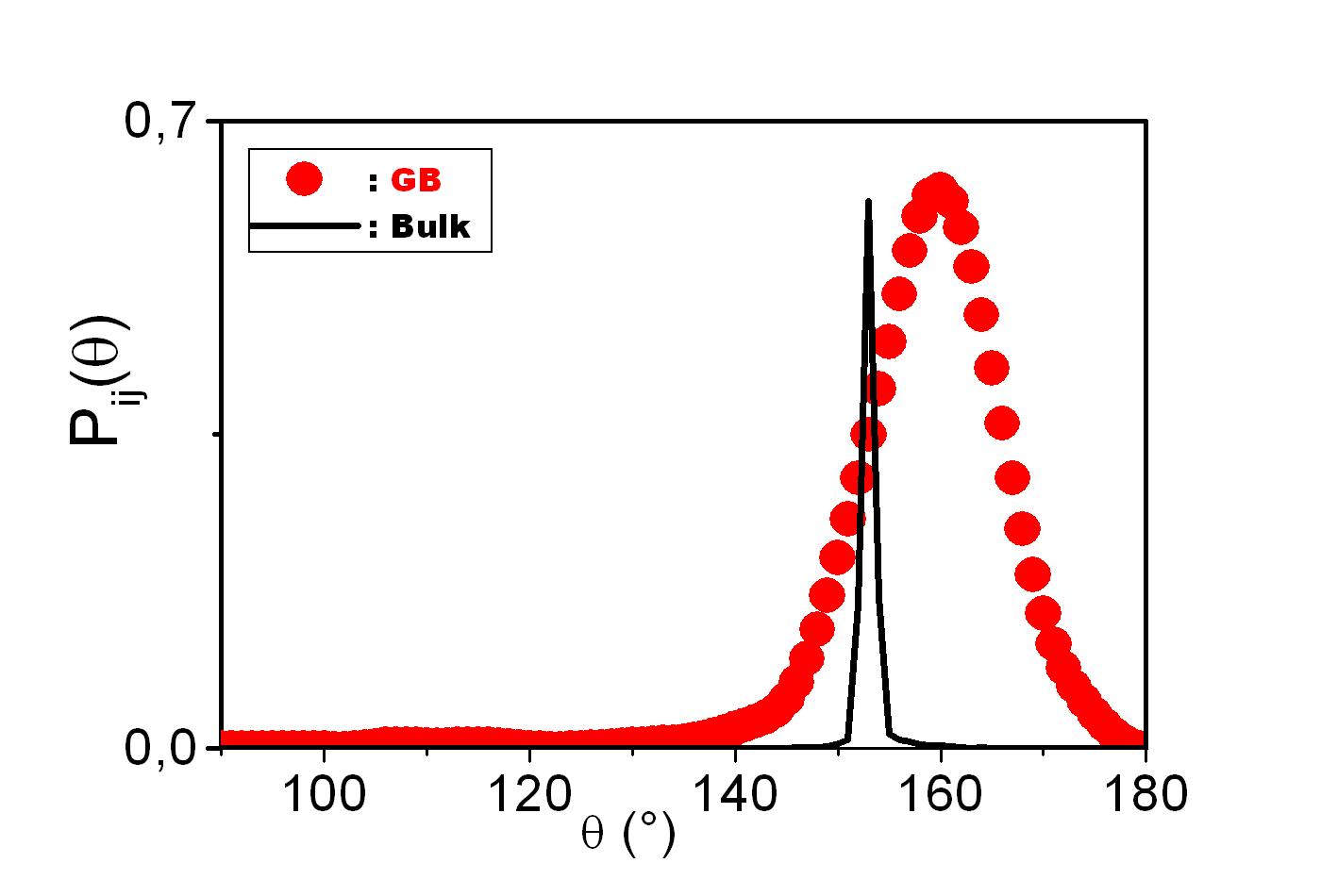}
    \caption{\textit{(a):  Coupling constant  as a function of the superexchange
    angle for iron fluoride.(b):  Angular distribution function for
    the superexchange angle.  GB is characterized by a large peak around
    160°.}}
    \label{Chap47}
\end{figure*}

Our sample model is chosen as 
 a double  grain boundary
constrained in between grains which consists of pure perfectly
crystalline $ FeF_3 $ with a size ranging from 8 to 12 nanometers as
experimentally measured in iron fluoride prepared by high energy ball
milling.  Those samples are obtained by Vorono\"i tessellation and
structurally relaxed with a scheme based on a modified Metropolis
algorithm
\cite{Fongang}.  It has been shown both experimentally and numerically
in such a system that the grain boundaries are disordered, 
but still remain  composed of
corner-sharing octahedral units with odd and even iron rings, thus 
leading  to
magnetic frustration.  
The macroscopic thermodynamic properties, such as the temperature
dependence of the magnetization, the specific heat and the 
magnetic susceptibility
for our system, are obtained from a Heisenberg-type Hamiltonian, which in
general contains several terms corresponding to different energy
contribution:  exchange, Zeeman, dipolar, anisotropy, magnetostatic,
magnetoelastic and thermal energy.  In our study, we only consider the
exchange energy for the first part of the work and then we add Zeeman
contribution for the study of hysteresis loops.  The Hamiltonian of the
system is thus
\begin{equation}
    H = \sum_{i=1}^N [-\frac{1}{2}\sum_{j\in V}J_{ij}\vec{S}_i\vec{S}_j
    -\mu_0 g_i\mu_B\vec{S}_i\vec{H}_{ext}]%
    \label{eqnHamil}
\end{equation}
$ V $ is the nearest neighborhood of site $ i $ , $ J_{ij} $ are the
exchange coupling constant, $ \vec{S_i} $ and $ \vec{S_j} $ are spins
corresponding to the $ i $ an $ j $ sites, $ \mu_b $ is the Bohr
magneton,$ g_i $ is the gyromagnetic ratio and $ H_{ext} $ is the
external magnetic field.\\
The first step of our work consisted in the determination of the coupling
constant $ J_{ij} $ which is related to the superexchange angle by the
relation
\begin{equation}
    J(\theta)=J_{180^\circ}\cos^2\theta + J_{90^\circ}\sin^2\theta
\end{equation}
which can be written in the form
\begin{equation}
    J(\theta)=J_{90^\circ} + (J_{180^\circ}-J_{90^\circ})\cos^2\theta%
    \label{eq53}
\end{equation}
where $ J_{90^\circ} $ and $ J_{180^\circ} $ are respectively 
the coupling constant corresponding 
to the superexchange angle of $ 90^\circ $ and $
180^\circ $.\\
From previous studies several pieces of informations are available for  r-$ FeF_3
$ :
\begin{itemize}
    \item
        The N\'eel temperature is approximatively $ 363 K $
        \cite{Greneche3};
    \item
        The blank angle (angle for which the constant coupling is zero)
        is $ 115^\circ $
        \cite{Lacorre};
    \item
        The superexchange angle in this phase is $ 153.15^\circ $
        \cite{Leblanc}.
\end{itemize}
Taking into account the fact 
that each iron atom has six F neighbors and that
octahedral units are regular in pure perfectly crystalline r-$ FeF_3 $, we can
use the relation giving the N\'eel temperature as a function of 
the  coupling constant 
to determine the value of $ J_{\theta_R} $ in units of
the Boltzmann constant $k_B$
\begin{equation}
    T_N = 1.45\frac{Z|J_{\theta_R}| S^2}{6k_B}%
    \label{EqnNeel}
\end{equation}
where $ Z $ is the coordination number and $ S $ the spin.  With $ Z=6 $
and $ S=1 $ $ J_{\theta_R} = -250.35  $ K \\
By combining relation~\ref{eq53} for the rhombohedral superexchange angle ($ \theta_R $) and
the blank angle ($ \theta_B $), we obtain the system
\begin{equation}
    \left\{
    \begin{array}{c}
        J(\theta_R)=J_{90^\circ}(1-\cos^2\theta_R) + J_{180^\circ}\cos^2\theta_R
        \\
        0 =J_{90^\circ}(1-\cos^2\theta_B) + J_{180^\circ}\cos^2\theta_B
    \end{array}
    \right\}
\end{equation}
The resolution of this system gives
\begin{equation}
    J_{90^\circ} = \frac{J_{\theta_R}\cos^2\theta_B}{\cos^2\theta_B-\cos^2\theta_R}
\end{equation}
and
\begin{equation}
    J_{180^\circ} = \frac{J(\theta_R)(\cos^2\theta_B -1)}{\cos^2\theta_B-\cos^2\theta_R}
\end{equation}
which gives $ J_{90^\circ}=72.42 K $ and $ J_{180^\circ} = -333.07 K $.\\
We can then deduce the relation giving the coupling constant as a
function of the superexchange angle.  This relation is represented in figure~%
\ref{Chap47}.  A comparison with the distribution of the superexchange angle
in  nanostructured $ FeF_3 $ obtained from previous studies
\cite{Fongang} implies that all interactions in the GB remain
antiferromagnetic.  This is observed in figure~\ref{Chap47} 
by the large peak centered
around $ 160^\circ $, all angles being greater than $ 140^\circ $. 
It is thus concluded that the frustration in the GB does only
originate from the cationic topology.
\begin{equation}
    J(\theta)=72.42 -405.49\cos^2\theta%
    \label{EqnJTheta}
\end{equation}

In
the first step of the simulations
we only considered the exchange contribution to the
magnetic Hamiltonian of the system.  We have then neglected 
other
phenomenological contributions such as 
dipolar, Zeeman or  anisotropy terms,  
in order to clarify the influence of the exchange coupling 
on the magnetic behavior. 
All samples, with sizes ranging
from 8 to 12 nanometers and different orientations of grains (tilted,
twisted etc) were first structurally relaxed. The energy given in
equation%
\ref{eqnHamil} is minimized by means of the Monte Carlo/Metropolis
simulated annealing procedure.
During the simulation,  Monte Carlo steps are applied on individual
magnetic moments ("spins")  with a random
walk while  solid angles are uniformly distributed over $ 4\pi $.  Starting
with a random spin configuration at a temperature $T_{\mbox{start}}$
much higher than $ T_N$
 (600K), the energy is minimized using the simulated annealing scheme with
a decreasing power law for temperature 
$T=T_{\mbox{start}} 0.97^i$ where $i$ is the step number, while
the thermodynamic quantities, such as magnetization, susceptibility and
specific heat, can be derived as a function of the temperature.  In each
simulation, the final temperature is  lower than 1K and the number of
Monte Carlo steps per spin (MCS) is $ 10^5 $.  \\

\section{Results} 
\begin{figure*}[!ht]
    \centering
    \includegraphics[scale=0.43]{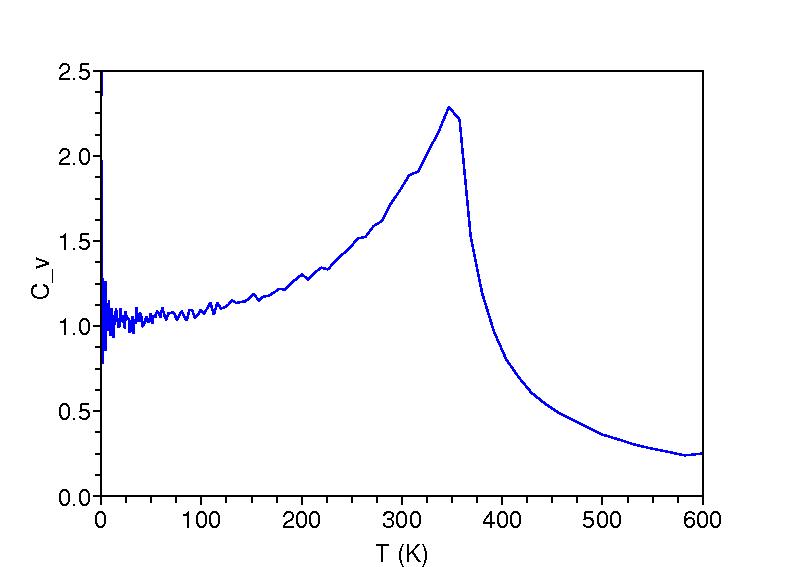}
    \includegraphics[scale=0.2]{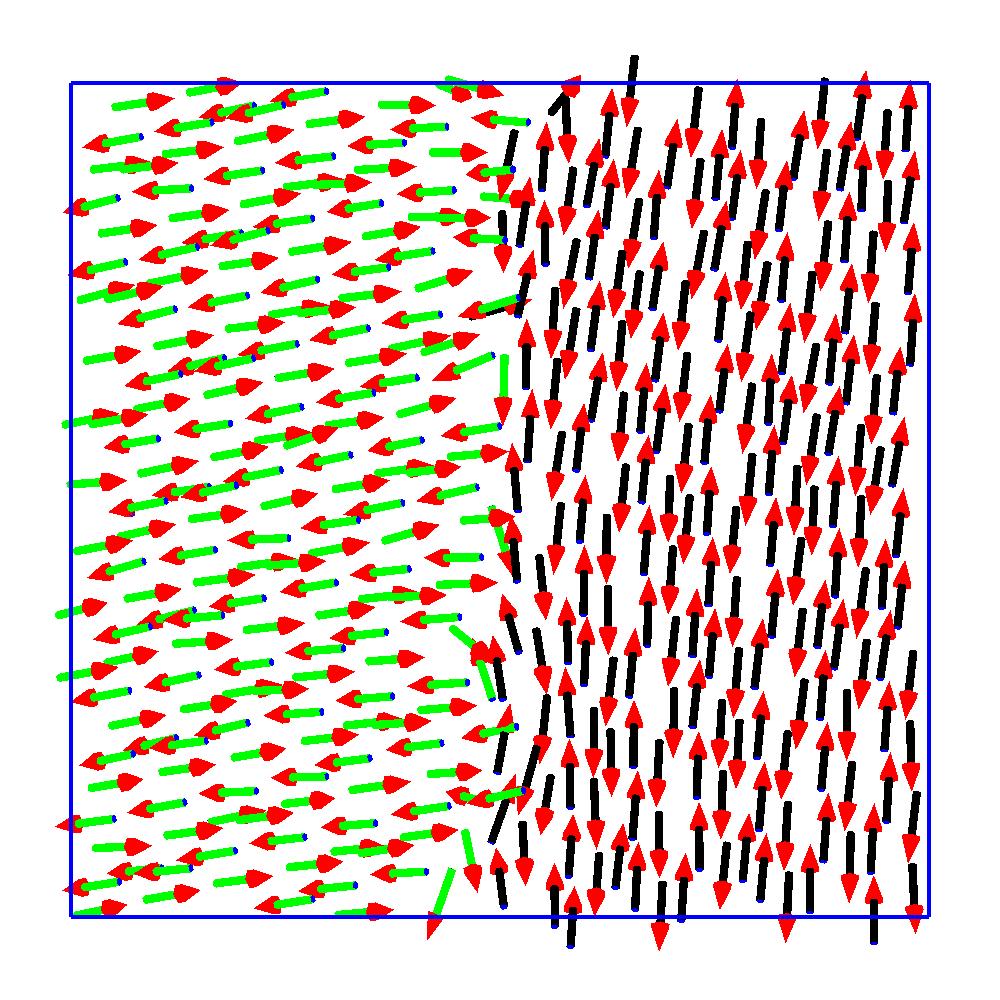}
    \caption{\textit{(a):  specific heat of the system as a function of
    temperature (N\'eel temperature is around 360K).(b):  Magnetic
    configuration after annealing:  each change in simulation conditions
    would give a different orientation of the total moment of the grain.}}
    \label{Chap47b}
\end{figure*}

We first determine the magnetization, susceptibility
and specific heat of the system as a function of temperature.  
Figure~\ref{Chap47b} represents the specific heat and one can observe a
transition around 360 K which is the N\'eel temperature of r-FeF3.  The
magnetization of the system is uniformly zero due to antiferromagnetic
behavior.  The magnetic configuration after annealing is composed of
two antiferromagnetic grains separated by a transition zone with
disordered spin structure as illustrated  on figure~
\ref{Chap47b}.  Each change in simulation conditions (number of MCS,
orientation of initial grains, final temperature, etc) leads to a
different orientation of magnetic moments of grains and then a different
coupling of GB moment.  \\

\begin{figure*}[!ht]
    \centering
    \includegraphics[scale=0.17]{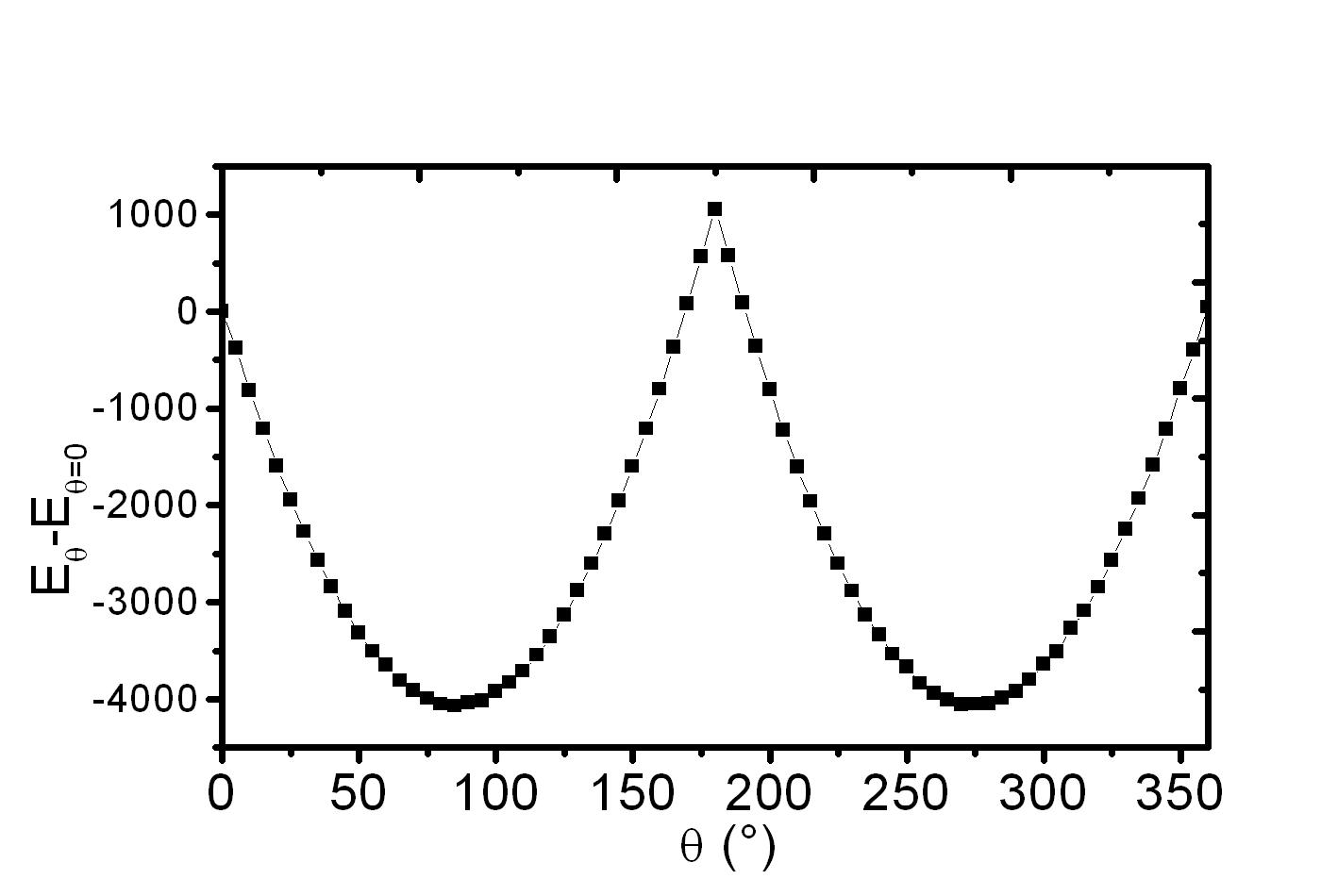}
    \includegraphics[scale=0.17]{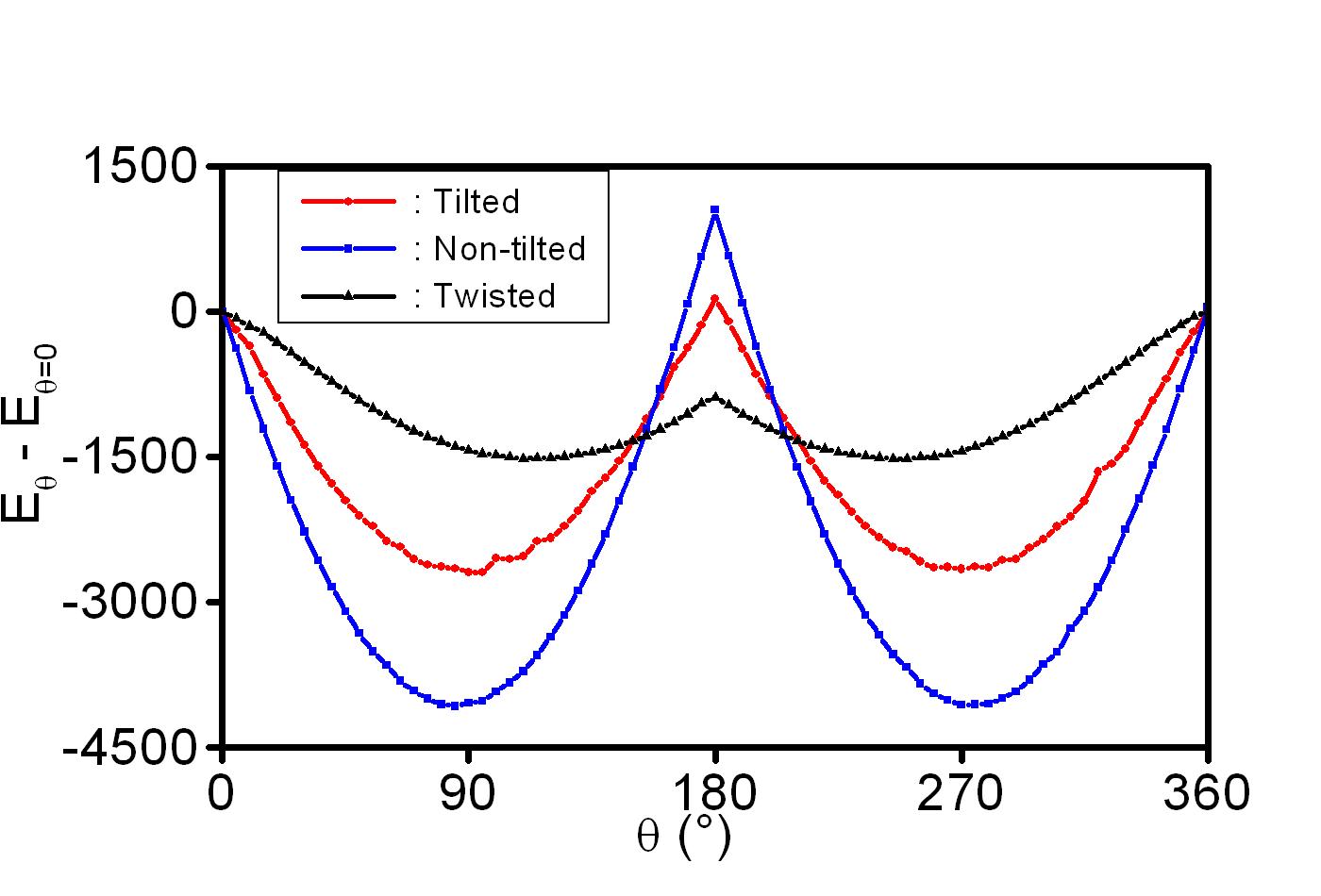}
    \caption{\textit{(a):Magnetic energy as a function of fixed angle
    boundary condition.  There are two magnetic states of
    lower energy around $ 90^\circ $ and $ 270^\circ $.  (b):  The same
    behavior is obtained for tilted, twisted and non-tilted system but
    with different angles.}}
    \label{Chap47c}
\end{figure*}

We can conclude that the system is not in the magnetic equilibrium
state.  This can be easily understood by considering the fact that the GB is
magnetically frustrated.  The present single spin flip we used is not
appropriate for finding the magnetic equilibrium state in a reasonable
time of simulation.  New conditions have to be applied to reach this magnetic
equilibrium state. A constraint can be added by considering the fact that there
is one orientation of the relative global moment of
 each grain for which the
magnetic energy is minimal.  We used fixed boundary conditions to
constrain magnetic moments of the two grains to form an angle $\theta$¸ and
then relaxed the interface.  \\
Figure~
\ref{Chap47c} represents the relative energy of the system for an angle
varying from $ 0^\circ $ to $ 360^\circ $.  We can clearly distinguish two
magnetic states with minimal energy around $ 90^\circ $ and $ 270^\circ $.

\begin{figure*}[!ht]
    \includegraphics[scale=0.2]{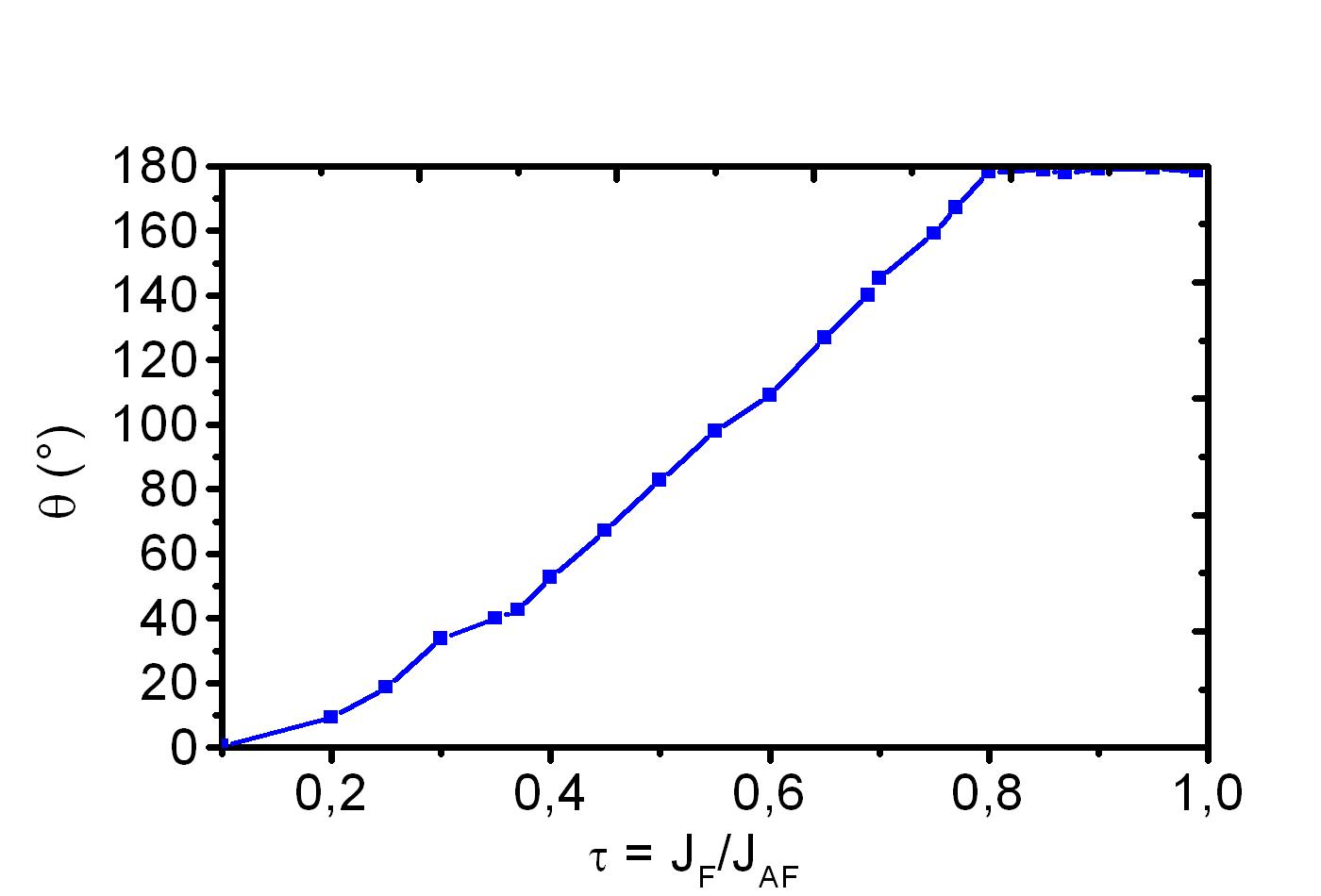}
    \caption{
    \label{Chap511Ang}\textit{Coupling angle between grains as a
    function of frustration rate.  Non collinear coupling appears for
    a rate comprised between 0.2 and 0.8.}}
    \label{Couplage}
\end{figure*}

The same behavior of the magnetic energy was obtained for different
grain orientations, but the angle corresponding to the minimal energy
was not the same.  Figure~
\ref{Chap47c} compares the same results obtained with non-tilted, tilted and
twisted samples :   it clearly appears  
that the angle corresponding to the minimum
energy depends on the initial relative crystallographic
orientation of grains.  It has recently  been
shown~\cite{Fongang}   that the only difference between those
samples is the rate of frustration in the GB, suggesting that the angle
corresponding to the minimum energy depends on the rate of frustration.\\
To verify this dependence, we first considered  a simple model consisting of a
cubic cluster with all interactions being antiferromagnetic, then we
changed some 
antiferromagnetic in ferromagnetic interactions at the
interface to create locally a bond frustration.
Frustration rate can be thus  defined as $ \tau = NJ_F/(NJ_{AF}+NJ_{F}) $,
where $ NJ_F $ and $ NJ_{AF} $ respectively represent the number of 
ferromagnetic and antiferromagnetic
bonds.  For each rate, the system is relaxed and
the angle between the moments of both sides of interface is measured: 
a certain rate of frustration gives rise to   a non
collinear orientation of the moments as evidenced on
figure~ %
\ref{Couplage}.
  
\begin{figure*}[!ht]
\begin{center}
\begin{minipage}[t]{0.33\textwidth}
    \includegraphics[scale=0.18]{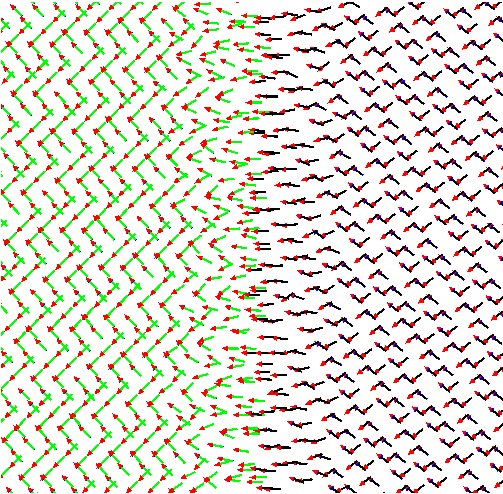}
 \end{minipage}
\begin{minipage}[t]{0.3\textwidth}
\includegraphics[scale=0.09]{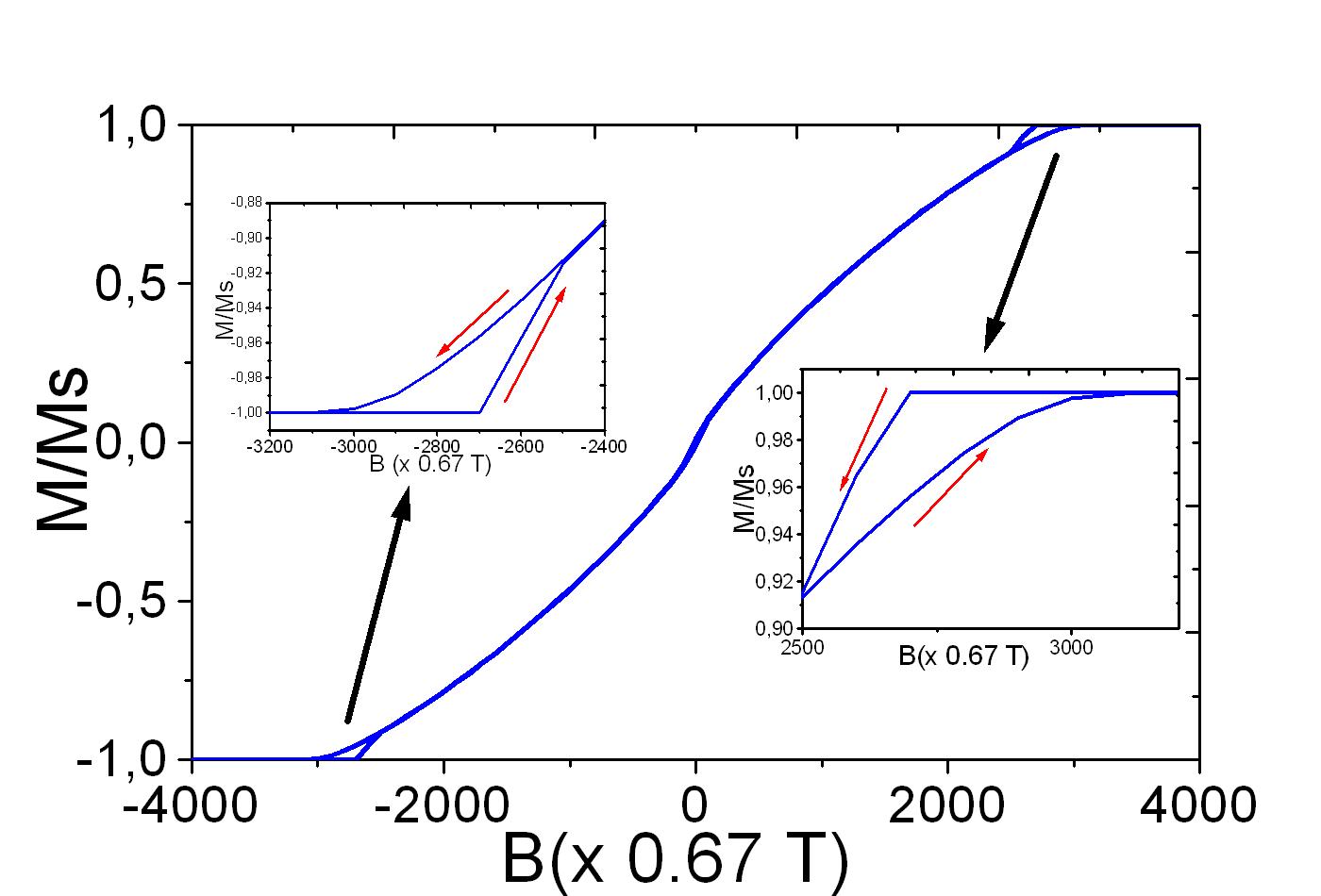}\\
 \end{minipage}
\begin{minipage}[t]{0.33\textwidth}
     \includegraphics[scale=0.18]{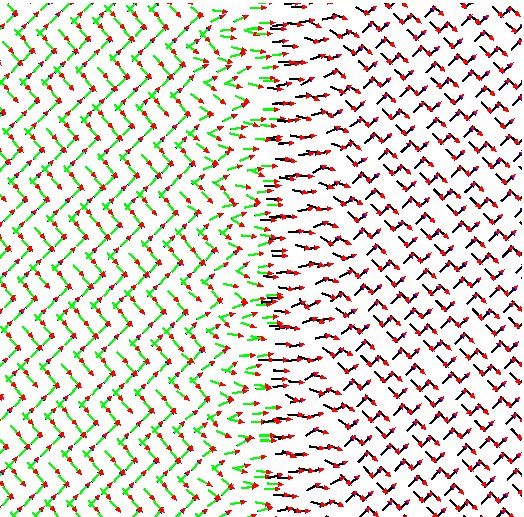}
 \end{minipage}
\end{center}
    \caption{\textit{Hysteresis loops and snapshots of the magnetic
        configuration during spin rotation for a 12nm size nanostructure.  As
    expected, there is no remanent magnetization but a non uniform spin
    rotation in the area in inset.  
     As the strength of the magnetic
    coupling far from the GB is greater than the one in the GB, magnetic moments
    in the GB are still in the direction of applied field after the
    spins far from the GB have rotated.}}%
    \label{Chap47d}
\end{figure*}

Therefore, we can assume that the non collinear orientation observed in
ferric fluoride nanostructures is due to the magnetic frustration of
the grain boundary. We conclude that the angle between the 
total momenta of the two grains depends on the rate of
frustration, which suggests the possibility to tune the final orientation of
the grains.
To achieve the study of this system, we have performed a simulation of
hysteresis loops.  It is  well established that  crystalline r-$ FeF_3 $ is
antiferromagnetic and that the GB behaves as a speromagnet giving rise
to zero magnetization : 
therefore we do not expect  any remanence on the hysteresis
loops.  The goal of the study is to analyze the influence of disorder in 
the GB on the spin rotation under a magnetic field.\\
The hysteresis loops have been computed starting from a saturated state achieved
after the application of a high enough field (2700 T) along the z-axis
(orthogonal to the GB plane) and decreasing the field in constant steps $ \delta h = 10 T $, during
which the magnetization was averaged over 5000 MC steps after
thermalization.  The temperature was sufficiently low ($ 10^{-3} $ K) to
avoid thermal fluctuations.  The result for a sample with a 12 nm
size is presented in figure~%
\ref{Chap47d}.  As expected, there is no remanent magnetization in
hysteresis loops, but  the spin rotation from a configuration
parallel to the magnetic field to an antiferromagnetic configuration is not
uniformly done:  the moments far away from the interface are 
decoupling first due to their strong coupling compared to those of the GB.
\section{Conclusion} We presented a Monte Carlo study of nanostructured
ferric fluoride
by coupling a previously established structural atomistic modeling to a
magnetic modeling. 
As experimentally evidenced, at the nanoscale, this
compound is composed of two parts:  A perfect antiferromagnetic
configuration in the grain and 
a disordered magnetic configuration (speromagnetic
like) at the grain boundary.  The present numeric
study has shown that the magnetic configuration at the
grain boundaries is governed by the cationic topology which
results from the disorientation of crystalline neighboring grains.
Consequently, the lowest magnetic energy of the total system
is strongly correlated to the rate of magnetic frustration in 
the GB, opening the possibility to tune the final
orientation of the magnetic moments of the grains by modulating
the frustration rate. 

From the careful simulation of hysteresis loops, the magnetization
reversal is not uniform under a magnetic field ; in addition
it is dependent on the strength of the magnetic coupling
which is related to the superexchange angle.  It is finally
important to emphasize that such an approach to numerically
model structural and magnetic structures can easily be extended
to ionic nanostructures, particularly to multilayers composed
of different oxide species in order to follow the dependence
of their properties on  their topological ordering,
the thickness of the interfaces and the mismatch between layers.

\bibliographystyle{model1-num-names}
\bibliography{publ}
\end{document}